\documentstyle[prl,aps,twocolumn,floats,graphicx,amsfonts,amssymb,%
amsmath]{revtex}
\begin{document}
\draft
\preprint{}
\title{Decay of Superflow Confined in Thin Torus: \\
A Realization of Tunneling Quantum Fields
}
\author{M. Nishida and S. Kurihara}
\address{
Department of Physics, Waseda University, 3-Okubo, Shinjuku-ku,
Tokyo 169-8555, Japan}
\date{\today}
\twocolumn[\hsize\textwidth\columnwidth\hsize\csname@twocolumnfalse%
\endcsname
\maketitle
\begin{abstract}
 The quantum nucleation of phase slips in neutral superfluids confined
 in a thin torus is investigated by means of the collective coordinate
 method. We have devised, with numerical justification, a certain
 collective coordinate to describe the quantum nucleation process of a
 phase slip. Considering the quantum fluctuation around the local
 minimum of the action, we calculate the effective mass of the phase
 slip. Due to the coherence of the condensate throughout the torus, the
 effective mass is proportional to the circumference $L$ of the torus,
 and the decay rate has a strong exponential $L$-dependence. 
\end{abstract}
]
\narrowtext

The circulation of superflow in a macroscopic torus container is
quantized. However, it has been pointed out that the quantized
circulation can decay via the thermal nucleation of phase slips in the
case of very narrow torus, or extremely thin superconducting
wires\cite{little67,langer67,mccumber70,mueller98}.
If the temperature is low enough, it is probable that the nucleation due 
to the macroscopic quantum tunneling become
dominant\cite{freire97,saito89}. 
There have been several experiments of superconducting wires which, in
fact, indicate the existence of a crossover between the thermal
activation behavior in the high temperature regime and the quantum
tunneling behavior in the low temperature regime\cite{giordano94}.
We expect that similar experiments on the neutral Bose gas systems
will become available, in the near future, with the trapped atomic
gases\cite{BEC}, or with the liquid $\mathrm{^4}$He in meso-porous
materials\cite{shirahama90}. The purpose of this Letter is to estimate
the probability of the quantum nucleation of phase slips in dilute Bose
gas systems.

At sufficiently low temperatures, a condensate in a dilute Bose gas
is described by a macroscopic wave function $\psi(\boldsymbol{r})$.
We consider a very thin torus container. The circumference $L$ is much
larger than the healing length $\xi=\frac{\hbar}{\sqrt{m\rho_0 g}}$,
where $m$ is the mass of an individual atom, $\rho_0$ is the density
of the condensate (without flow), and $g=\frac{4\pi\hbar^2a}{m}$ is the
interatomic interaction, with $a$ being the s-wave scattering
length. The cross-sectional area $\sigma$ is much smaller than
$\xi^2$. We assume that the spatial dependence of $\psi$ in the cross
section can be neglected by using the effective number density. We will
discuss this point later. Taking $x$ axis along the circumference of the 
container, the action $S[\psi]$ and the free energy $F[\psi]$ have the
Gross-Pitaevskii (GP) form\cite{gp}:
\begin{equation}
 S[\psi] =
  \int dt\,
  \left\{\sigma\int^{\frac{L}{2}}_{-\frac{L}{2}}dx
   \,\bar{\psi}\left(\text{i}\hbar\frac{\partial}{\partial t}\right)\psi
   - F[\psi]\right\},
 \label{eq:action}
\end{equation}
\begin{equation}
 F[\psi] = \sigma\int^{\frac{L}{2}}_{-\frac{L}{2}}dx\,
  \left\{\frac{\hbar^2}{2m}\left|\frac{\partial\psi}{\partial x}\right|^2
   + \frac{g}{2}\left(|\psi|^2-\frac{\mu}{g}\right)^2
	\right\},
 \label{eq:FreeEnergy}
\end{equation}
where $\mu \sim \rho_0 g$ is the chemical potential.

If we assume that the effective inertial mass of the phase slip is not
greatly dependent on the tunneling path, the most probable tunneling
path will be through the valley, passing through two local minima and
one saddle point in the functional space of $\psi$.
The wave function $\psi$ is determined so as to minimize the free energy
functional at each stage of the tunneling process.
We calculated the minimum value of the GP free energy numerically with
the constraint condition that the average velocity of the superflow had
each constant value.
We adopt the units in which the unit of length is the healing length
$\xi$, the unit of velocity is the sound velocity $c=\sqrt{\frac{\rho_0
g}{m}}$, and the unit of number density is the density of the condensate 
$\rho_0$. A dimensionless parameter, $\alpha\equiv\rho_0\sigma\xi$,
determines the importance of quantum fluctuations\cite{freire97}. We
assume $\alpha \gg 1$, and use WKB approximation. Considering
$\sigma\ll\xi^2$, $\alpha \ll \rho_0 \xi^3 \propto \rho_0^{-1/2}$, so
$\alpha$ is large when the density is low.  We
include the following term into the free energy Eq.\ (\ref{eq:FreeEnergy}):
\begin{equation}
\lambda\left[\int_{-\frac{L}{2}}^{\frac{L}{2}}dx
\rho(x)\left\{\frac{d\theta(x)}{dx}-v_{\text{s}}\right\}\right]^2,
\end{equation}
where $\rho(x)$ and $\theta(x)$ are the density and the phase of the
condensate respectively, $v_{\text{s}}$ is the constraint value of the
superflow velocity, and $\lambda$ is supposed to take sufficiently large
value which is determined by the condition that the difference between
the constraint value $v_{\text{s}}$ and the average velocity of the
obtained solution be smaller than, say, $1.0\times 10^{-6}c$. We had
$L=20\,\xi$ in a mesh of 100 points, and used the simulated annealing
method\cite{press86}. Because there are a lot of local minima which
correspond to multiple phase slip solutions, it is difficult to obtain
true minimum values with other methods. In order to avoid being trapped
in these local minima, we introduce random noises which correspond to
thermal fluctuations at given temperature. The calculation proceeds
gradually from high temperature region to low temperature region, and
continues until equilibrium values are reached. In our calculation, the
final temperature is $1.5\times 10^{-10}\,\rho_0 g$, and the expected
error of the free energy is $O\left(10^{-4}\rho_0 g\right)$.

\begin{figure}[htpb]
\begin{center}
\includegraphics[keepaspectratio=true, height=.37\textwidth, angle=270]%
{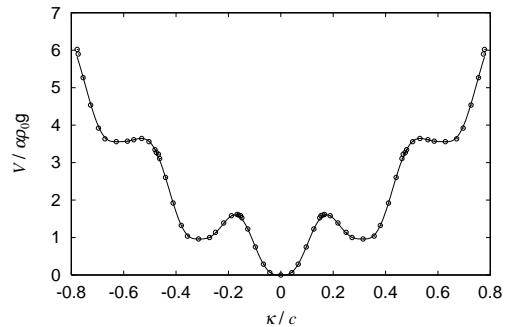}
\caption{Potential energy for the collective coordinate $\kappa$.
 The points are the minimum values of the GP free energy which were
 calculated numerically. The solid line denotes the value of the
 potential energy Eq.\  (\ref{eq:potential}) which is calculated using
 the interpolation wave function Eq.\ (\ref{eq:PhaseSlip}). }
\label{fig:potential}
\end{center}
\end{figure}

Extrema of the GP free energy functional are solutions to the
time-independent GP equation:
\begin{equation}
 -\frac{1}{2}\frac{\partial^2\psi}{\partial x^2}
 + \left(|\psi|^2-1\right)\psi = 0.
\end{equation}
There are two kinds of solutions to this
equation\cite{mccumber70}. The first kind of solution corresponds to a
free-energy local minimum:
\begin{equation}
 \psi_{\text{m}} = \text{i}\sqrt{1-\frac{\ \kappa^2}{2}} e^{\text{i}\kappa x},
 \label{eq:minimum}
\end{equation}
where $\kappa$ is the phase gradient or, in other words, the superflow
velocity.
The second kind corresponds to a free-energy saddle point:
\begin{equation}
 \psi_{\text{s}} = e^{\text{i}\kappa x}
  \left\{\sqrt{1-\tfrac{3}{2}\kappa^2}
   \tanh\left(\sqrt{1-\tfrac{3}{2}\kappa^2}x\right)
   -\text{i}\kappa\right\}.
 \label{eq:saddle}
\end{equation}
The phase gradient of $\psi_{\text{s}}$ approaches $\kappa$
asymptotically as we go away from the phase slip center
($x=0$). Hereafter, we call $\kappa$ ``asymptotic velocity''.
If we can find the function which interpolates these two kinds of
solutions smoothly, and fits the numerical results, we may use it as the 
wave function which describes the condensate during the whole nucleation
process. Considering the periodic boundary condition, we have found the
following interpolation function:
\begin{eqnarray}
 \psi_{\text{i}} & = & \sqrt{1-\frac{\ \kappa^2}{2}} e^{\text{i}\kappa x}
 \nonumber \\
 & & \times \left\{\sin\frac{q}{2}\tanh\left(\sqrt{1-\frac{\ \kappa^2}{2}}
	     \sin\frac{q}{2}\,x\right)+\text{i}\cos\frac{q}{2}\right\}.
  \label{eq:PhaseSlip}
\end{eqnarray}
We must determine the parameter $q$ so that $\psi_{\text{i}}$
satisfies the periodic boundary condition. If $qL\gg 1$, we can take the
form, 
\begin{equation}
 q=\kappa L-2\pi n, 
 \label{eq:q}
\end{equation}
where $n$ is the winding number of the phase around the circumference of
the container $(n=0,\pm 1,\pm 2,\ldots)$. The parameter $q$ expresses
the phase difference between two sides of the phase slip
center. Hereafter, we call it ``phase jump''. We can adopt either 
asymptotic velocity or phase jump as an appropriate collective
coordinate to describe the nucleation process.
Substituting Eq.\ (\ref{eq:PhaseSlip}) into Eq.\ (\ref{eq:FreeEnergy}),
the potential energy for the asymptotic velocity $\kappa$ is obtained as
follows:
\begin{eqnarray}
 V(\kappa)
 &=& \frac{4}{3}\left(1-\frac{\ \kappa^2}{2}\right)
 ^{\frac{3}{2}}\sin^3\frac{q}{2} \nonumber \\
 &-& 2\kappa\left(1-\frac{\ \kappa^2}{2}\right)\sin\frac{q}{2}
  \cos\frac{q}{2}
 +\frac{L}{2}\left(\kappa^2-\frac{\ \kappa^4}{4}\right).
 \label{eq:potential} 
\end{eqnarray}
Using $\kappa$ as a fitting parameter, we fit the amplitude and phase of 
the wave function Eq.\ (\ref{eq:PhaseSlip}) with those which are given
by the numerical minimization. The phase jump changes about $L(=20)\%$
when $\kappa$ changes $1\%$ (see Eq.\ (\ref{eq:q})), so the shape
of the phase is very sensitive to the value of $\kappa$.
Fig.\ \ref{fig:potential} shows that Eq.\ (\ref{eq:potential}) is in
good agreement with the numerical results.

We then consider the effective inertial mass for the motion of the
collective coordinate.
Because of the translational invariance, the total momentum must be
conserved between the initial state and the final state.
If there were no contributions from the noncondensed component, the
condensate would not lose the momentum and stay at the local minimum.
However, quantum fluctuations of $\psi$ exist around the distorted 
condensate wave function $\psi_{\text{i}}$, and these fluctuations will
play the role of a noncondensed component.
Therefore, the condensate can oscillate around the local minimum and
tunnel to another local minimum by exchanging the momentum with
quantum fluctuations, even at zero temperature.
The frequency of the zero-point oscillation of the collective
coordinates around the local minimum will have the same value of the
collective mode whose wave length is $L$. 
Because we have already determined the potential energy, Eq.\
(\ref{eq:potential}), we can estimate the effective inertial mass from
the frequency of the collective mode and the curvature of the potential
energy for $\kappa$ or $q$ at the local minimum. In this procedure, we
assume that the inertial mass is not greatly dependent on the collective 
coordinate, and approximate it to the value at local minimum.
The coupling with the collective modes also suggests that there are the
effects of dissipation, which will suppress the tunneling
probability\cite{caldeira81}. However, we will not take into account
these effects, but concentrate our interests on the bounce motion of
$\kappa$ or $q$. In order to clarify these two points, we would have to
consider quantum fluctuations explicitly and integrate them out, thus
requiring further study. 

In order to estimate the frequency of the collective mode, we consider
the fluctuation around the local minimum solution, Eq.\
(\ref{eq:minimum}).
We substitute
$\psi=\psi_{\text{m}}+(u+\text{i}v)\text{e}^{\text{i}\kappa x}$ into the 
action, Eq.\ (\ref{eq:action}), and retain the terms up to the second
order of $u,v$, then we have
\begin{eqnarray}
 S[u,v] & = & \hbar\alpha\int\!dt
  \left[-V(\kappa)+\int\!dx \left\{2iv\left(\frac{\partial}{\partial t}
   \sqrt{1-\frac{\ \kappa^2}{2}}\right) \right.\right. \nonumber \\
 & & + \left.\left. \frac{1}{2}
 \left(\begin{array}{cc}
  u & v
       \end{array}
  \right)
  \boldsymbol{G}^{-1}
  \left(\begin{array}{c}
   u \\
   v
	\end{array}
   \right)\right\}\right],
\end{eqnarray}
where $\boldsymbol{G}$ is the Green's function for $u,v$, whose Fourier
component is given by
\begin{eqnarray}
 \boldsymbol{G} & = &
 \left\{\frac{\ k^2}{2}\left(\frac{\ k^2}{2}
                     +2\left(1-\frac{\ \kappa^2}{2}\right)\right)
                     -(\kappa k-\omega)^2 \right\}^{-1} \nonumber \\
 & & \times \frac{1}{2}\left(\begin{array}{cc}
        -\frac{\ k^2}{2}-2\left(1-\frac{\ \kappa^2}{2}\right) &
        \text{i}(\kappa k-\omega) \\
        -\text{i}(\kappa k-\omega) &
        -\frac{\ k^2}{2}
       \end{array}
 \right).
\end{eqnarray}
$\boldsymbol{G}$ has the poles which correspond to the Bogoliubov modes
in the flowing medium. We can derive these poles by Galilei
transformation from the co-flowing frame to the lab frame, noting that
the density is now $1-\frac{\ \kappa^2}{2}$. From these poles, we
estimate the effective angular frequency of the oscillation of $\kappa$
by the following relation:
\begin{eqnarray}
 \omega & & _{\text{eff}}^2 - \omega^2 \nonumber \\
 & & = \left[\frac{k^2}{2}\left\{\frac{k^2}{2}
	+2\left(1-\frac{\kappa^2}{2}\right)\right\}
		       -(\kappa k-\omega)^2\right]\bigg|_{k=\frac{2\pi}
  {L}}.
\end{eqnarray}
Around the $n$th local minimum, whose velocity is $\kappa_n=\frac{2\pi
n}{L}$, the effective angular frequency becomes
\begin{equation}
 \omega_{\text{eff}}^2(n)=\left(\frac{2\pi}{L}\right)^2
 \left\{1-\frac{3}{2}\kappa_n^2+\left(\frac{\pi}{L}\right)^2
 \right\}.
 \label{eq:frequency}
\end{equation}

When q is so small that we cannot assume $qL\gg 1$, Eq.\
(\ref{eq:q}) and Eq.\ (\ref{eq:potential}) become inaccurate and we must
determine the relationship between $q$ and $\kappa$ with direct
consideration of the periodic boundary condition. After doing so, the
potential curvature at the $n$th local minimum becomes 
\begin{equation}
 V''(\kappa_n)=L\left(1-\frac{3}{2}\kappa_n^2\right),
 \label{eq:curvature}
\end{equation}
where the prime denotes differentiation with respect to $\kappa$.

From Eq.\ (\ref{eq:frequency}) and Eq.\ (\ref{eq:curvature}), the
effective inertial mass for $\kappa$ is given by
\begin{eqnarray}
 M_{\kappa}(n) & = & \frac{V''(\kappa_n)}{\omega_{\text{eff}}^2(n)}
               =  \frac{L^3}{(2\pi)^2}
	       \left(\frac{1-\frac{3}{2}\kappa_n^2}
		{1-\frac{3}{2}\kappa_n^2
		+\left(\frac{\pi}{L}\right)^2}\right)
              \nonumber \\
              &\sim& \frac{L^3}{(2\pi)^2}
	       \quad (L\gg 1).
\end{eqnarray}

We now consider the tunneling problem from $n$th local minimum to the
$(n-1)$th local minimum. The tunneling rate $\Gamma$ is given by
$\Gamma=A e^{-S_{\text{B}} / \hbar}$, where $S_{\text{B}}$ is the bounce
action, which is the Euclidean action for the bounce motion of
$\kappa$ \cite{coleman77}:
\begin{equation}
  S_{\text{B}}=\hbar\alpha\int d\tau
   \left\{\frac{M_{\kappa}(n)}{2}\dot{\kappa}^2
   + V(\kappa)-V(\kappa_n)\right\}.
\end{equation}
Because $\kappa$ contains either the winding number $n$ or the system
length $L$, it is more convenient to use the phase jump $q$ for
the collective coordinate when we consider the dependence of the
tunneling rate $\Gamma$ on the superflow velocity $\kappa_n$ or the
system length $L$. 
Rewriting the bounce action $S_{\text{B}}$ by using the phase jump
$q=(\kappa-\kappa_{n-1})L$, we obtain
\begin{equation}
 S_{\text{B}} = 2\hbar\alpha\int_{q_0}^{2\pi}\,dq \sqrt{2M_q(n)V_n(q)},
 \label{eq:bounce}
\end{equation}
where $q_0$ is the classical turning point of $-V_n(q)$, 
\begin{equation}
 M_q(n) = \frac{M_{\kappa}(n)}{L^2}
 \sim \frac{L}{(2\pi)^2}
 \quad (L\gg 1),
 \label{eq:mass}
\end{equation}
\begin{multline}
 V_n(q) = V(q/L+\kappa_{n-1})-V(\kappa_n) \\
       \sim \frac{4}{3}\left(1-\frac{\kappa_n^2}{2}\right)
	^{\frac{3}{2}}
	\sin^3\frac{q}{2}-\kappa_n
	\left(1-\frac{\kappa_n^2}{2}\right)\sin q \\
         (L\gg 1).
\end{multline}
We thus find that the effective inertial mass for the nucleation of a
phase slip is proportional to the system length $L$. It
basically comes from the fact that the potential curvature is
proportional to the system length, (see Eq.\ (\ref{eq:curvature})). 
When a phase slip nucleates, the whole wave function of the condensate
changes its configuration due to its overall coherence, so the
free energy change becomes larger as the system circumference becomes
longer. Therefore the coherence of the condensate wave function is the
origin of the $L$-dependence of the inertial mass.

We can calculate the bounce action
analytically in the zero-current limit $(\kappa_n \to 0)$:
\begin{eqnarray}
 S_0 & = & \frac{2\sqrt{2}\hbar\alpha\sqrt{L}}{\sqrt{3}\pi}
  \int_{0}^{2\pi} \, dq
  \sin^{\frac{3}{2}}\frac{q}{2} \nonumber \\
 & \sim & 1.82\,\hbar\alpha\sqrt{L}.
\end{eqnarray}

\begin{figure}[htpb]
\begin{center}
\includegraphics[keepaspectratio=true, height=.37\textwidth, angle=270]%
{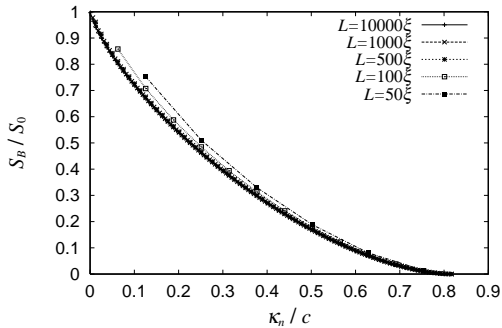}
\caption{Bounce actions normalized by $S_0$, which are calculated for
 the five different samples with the circumferences of
 $50-10000\,\xi$.}
\label{fig:action}
\end{center}
\end{figure}

Fig.\ \ref{fig:action} shows the bounce actions normalized by $S_0$
in the five different cases with the circumferences of
$50-10000\,\xi$. In all cases, the ratio $S_{\text{B}}/S_0$ goes to zero
at the common  velocity $v_{\text{c}}\sim\sqrt{2/3}\,c$. 
This velocity corresponds to the critical velocity of the superflow. If
the asymptotic velocity $\kappa$ goes beyond $v_{\text{c}}$, the saddle
point solution to the time-independent GP equation, Eq.\
(\ref{eq:saddle}), will vanish. No stable local minima exist for
velocity $\kappa_n>v_{\text{c}}$. The value of $v_{\text{c}}$ agrees
with the one which is derived by Landau's criterion. As $\kappa_n$
approaches the value of $v_{\text{c}}$, the bounce action becomes
extremely small. In the case of $L=100\,\xi, n=12$, the normalized
bounce action $S_{\text{B}}/S_0$ becomes $0.012$.

We see from Fig.\ \ref{fig:action} that there is no sizable
$L$-dependence for $L \gg 1$ under consideration. If $L$ is larger than
$100\,\xi$, all the data seem to be on the same line. Therefore we
conclude that the $L$-dependence mainly comes from the effective
inertial mass, and the bounce action is proportional to
$L^{\frac{1}{2}}$. This situation is completely different from that in
the thermal nucleation processes. In the thermal nucleation processes,
the decay rate is estimated by the Arrhenius formula $\Gamma \propto
e^{-\delta F/k_{\text{B}} T}$, which is determined only by the barrier
height $\delta F$, so there is no dependence of the decay rate on the
system length\cite{little67,langer67,mccumber70,mueller98}.  In the
quantum nucleation, the coherence of the condensate wave function
becomes important, and the decay rate has the strong $L$-dependence:
\begin{equation}
 \Gamma \propto \exp\{-1.82\,\alpha \sqrt{L}\,f(\kappa_n)\},
\end{equation}
where $f(\kappa_n)$ is the function which depends only on the superflow
velocity, and whose shape is shown in Fig.\ \ref{fig:action}. 

We now discuss the spatial dependence of $\psi$ in the cross section. 
If we consider a complete set of radial eigenfunctions, the energy
scale of the $\nu$th eigenstate will become $\frac{\hbar^2
\nu^2}{m\sigma}$. Because it is much larger than $\mu \sim
\frac{\hbar^2}{m\xi^2}$ in our model when $\nu \neq 0$, the
$\nu=0$ channel will become dominant\cite{mueller98}. Besides, the
energy scale of the potential barrier is $\alpha\mu \sim
\frac{\hbar^2\sigma\rho_0}{m\xi}$, which corresponds to the
energy of $\frac{\rho_0\sigma^2}{\xi\nu^2}$ particles excited into the
$\nu$th channel. The ratio to the total number is $\frac{\sigma}{\xi L
\nu^2} \ll 1$ when $\nu \neq 0$. These estimations seem to indicate that
we can safely neglect the spatial dependence in the cross
section. However, there are other effects to be considered: For example,
the nonlinearity of GP equation. It will cause couplings between the
current direction and the radial direction. If we take this into
account, we may find a more appropriate tunneling path. Moreover, when
the condensate is squeezed in a region whose size is smaller than the
healing length, the phase of the condensate will fluctuate
intensively. Because phase fluctuations $\left<(\delta\theta)^2\right>$
will effectively suppress the amplitude of the wave function by the
factor $1-\frac{1}{2}\left<(\delta\theta)^2\right>
+O\left((\delta\theta)^4\right)$.
This means that the effective number density of the condensate, and thus 
the parameter $\alpha$, which gives the scale of bounce action, will
become smaller than the results of the present calculation.
Therefore, it is probable that our result of tunneling rate is an
underestimate. The analysis of these effects is left for the future
work.

In conclusion, we have studied the quantum nucleation of phase slips in 
torus shaped superfluids by means of the collective coordinate method. 
We have found a wave function which interpolates the local minimum and
saddle point solutions to the time-independent GP equation, and this
fits our numerical results.  We have determined the appropriate
collective coordinate which describes the whole nucleation process,
and obtained the analytic form of the potential barrier. We have shown
that the effective inertial mass is proportional to the system
length $L$ due to the coherence of the condensate. Because the length
dependence of the bounce action mainly comes from the mass, we
have concluded that the bounce action is proportional to
$L^\frac{1}{2}$.  
\\ \\
We thank T. Inoue, I. Tomita, L. Mowrey, and T. Minoguchi for useful
discussion. This work is supported by Grant-in-Aid for Scientific
Research from the Ministry of Education, Science, Sports, and Culture of 
Japan.


%
%
\end{document}